\documentclass{iopjournal}
\usepackage{amsmath, amssymb, mathtools, mathrsfs}
\usepackage[colorlinks=true]{hyperref}
\hypersetup{colorlinks,linkcolor={red},citecolor={blue},urlcolor={blue}}  
\usepackage{color}
\usepackage{graphicx}
\usepackage[normalem]{ulem}
\usepackage{mathtools}
\usepackage{stmaryrd}
\usepackage{float}
\usepackage{textalpha}

\newcommand{\ev}[1]{{\langle{#1}\rangle}}

\begin{document}

\articletype{Paper} 

\title{From percolation transition to Anderson localization in one-dimensional speckle potentials}

\author{
Margaux Vrech$^1$, 
Jan Major$^2$,
Dominique Delande$^2$,
Marcel Filoche$^1$ and 
Nicolas Cherroret$^{2,*}$
}

\affil{$^1$Institut Langevin, ESPCI Paris, Universit\'e PSL, CNRS, 75005 Paris, France}

\affil{$^2$Laboratoire Kastler Brossel, Sorbonne Universit\'{e}, CNRS, ENS-PSL Research University, 
Coll\`{e}ge de France; 4 Place Jussieu, 75005 Paris, France}

\affil{$^*$Author to whom any correspondence should be addressed.}

\email{nicolas.cherroret@lkb.upmc.fr}


\begin{abstract}
\vspace{0.3cm}
\small
Classical particles in random potentials typically experience a percolation phase transition, being trapped in clusters of mean size $\chi$ that diverges algebraically at a percolation threshold. In contrast, quantum transport in random potentials is controlled by the Anderson localization length, which shows no distinct feature at this classical critical point. Here, we present a comprehensive theoretical analysis of the semi-classical crossover between these two regimes by studying particle propagation in a one-dimensional, red speckle potential, which hosts a percolation transition at its upper bound. As the system deviates from the classical limit, we find that the algebraic divergence of $\chi$ continuously connects to a smooth yet non-analytic increase of the localization length. We characterize this behavior both numerically and theoretically using a semi-classical approach. In this crossover regime, the correlated and non-Gaussian nature of the speckle potential becomes essential, causing the standard Dorokhov-Mello-Pereyra-Kumar (DPMK) description for uncorrelated disorder to break down. Instead, we predict the emergence of a bimodal transmission distribution, a behavior normally absent in one dimension, which we capture within our semi-classical analysis. Deep in the quantum regime, the DMPK framework is recovered and the universal features of Anderson localization reappear.
\end{abstract}

\section{Introduction}
\vspace{0.3cm}

In quantum disordered systems, subtle interference effects arising from multiple scattering events on random potential fluctuations are responsible for Anderson localization \cite{Anderson1958, Lee1985}. Over the past decades, this phenomenon has attracted renewed attention thanks to major advances in quantum simulators—most notably ultracold atomic gases, precisely controlled using laser light and other external  fields employed to confine the atoms or to tailor various types of disordered potentials~\cite{Aspect2009}. This high level of control has made quantum gases an exceptionally versatile platform for exploring Anderson physics, from weakly interacting systems~\cite{Cherroret2021, Shapiro2012}, to setups involving artificial gauge fields~\cite{Hainaut2018, An2018} or correlated regimes where many-body localization emerges \cite{Abanin2019}. In dilute gases, these developments have led to seminal experimental observations of Anderson localization in one-dimensional (1D) \cite{Billy2008, Roati2008} and three-dimensional \cite{Jendrzejewski2012, Semeghini2015} configurations, along with detailed investigations of its quantum-chaotic analogue in kicked-rotor models \cite{Moore1995, Chabe2008}.

A distinctive feature of experiments on disordered quantum gases is the use of random potentials exhibiting significant spatial \emph{correlations} and, in general, non-Gaussian statistics. Among these, optical speckle potentials have been widely employed, and their specific statistical properties have been shown to give rise to peculiar features of Anderson localization, such as effective mobility edges in one dimension \cite{Lugan2009, Izrailev2012} or non-monotonic energy dependence of the localization length \cite{Piraud2013}. Accounting for disorder correlations has also proven crucial for accurately characterizing the localization dynamics of cold atoms in position \cite{Sanchez-Palencia2007, Billy2008, Skipetrov2008} and momentum~\cite{Cherroret2012, Karpiuk2012, Ghosh2014, Lee2014} space, as well as for proper estimations of the localization length \cite{Sanchez-Palencia2007, Kuhn2007}, and of the mobility edge in three dimensions \cite{Yedjour2010, Delande2014, Pasek2017}. More broadly, correlations may profoundly modify the localization scenario, as exemplified by kicked-rotor systems where they give rise to a diffusive regime in one dimension \cite{Casati1990, Moore1995}, or induce genuine mobility edges in appropriately designed quasi-periodic systems (of infinite-range correlations) \cite{Biddle2010, Ganeshan2015, Wang2020}. In the context of optics, light in correlated media has also been shown to exhibit a variety of transport regimes \cite{Vynck2023}.

Another interesting feature of disordered speckle potentials is their percolation transition. In classical physics, percolation theory investigates the conditions under which randomly distributed objects form connected structures (``clusters'') that enable transport over arbitrarily long distances~\cite{Kirkpatrick1973, Isichenko1992, Saberi2015}. The critical point at which such large-scale connectivity emerges is known as the percolation threshold; it marks a geometric phase transition characterized by nontrivial critical behavior. In continuous speckle potentials, percolation thresholds have been identified in one, two, and three dimensions, corresponding to the energy level above which a classical particle can travel infinitely far along potential valleys \cite{Pezze2011, Jendrzejewski2012, Morong2015}. At the quantum level, however, the transport scenario changes drastically. In one and two dimensions, quantum particles are always localized, whereas in three dimensions an Anderson metal-insulator transition occurs \cite{Abrahams1979}---but at a point that differs markedly from the classical percolation threshold \cite{Jendrzejewski2012, Filoche2024}. Quantum mechanically, no direct trace of a critical percolation transition therefore appears to survive. Still, as one approaches the classical limit the Anderson and percolation problems should coincide, and the percolation transition re-emerge. How this connection establishes is the question we address in the present work, focusing on particle transport in a  1D speckle disorder---a simple and experimentally relevant continuous model which exhibits a percolation transition.

In detail, in this work we theoretically characterize the localization length near the percolation threshold as the system approaches its classical limit. This limit is governed by an effective Planck constant, $\hbar_{\mathrm{eff}}$, measuring the ratio between the de Broglie wavelength and the disorder correlation length.
As this ratio increases from zero, we find that the percolation transition smoothens, with the localization length crossing over from a strict divergence at the percolation threshold when $\hbar_{\mathrm{eff}} = 0$, to a continuous—though extremely steep— growth as $\hbar_{\mathrm{eff}} \ll 1$ becomes nonzero.
We describe this crossover both numerically and theoretically, by combining advanced properties of speckle statistics with semi-classical tools \cite{Prat2016, Trappe2015}. By analyzing the localization length and the statistical distribution of the transmission, we further show that near the percolation threshold the standard 1D Anderson localization theory for uncorrelated Gaussian disorder (Fokker-Planck/Dorokhov-Mello-Pereyra-Kumar or DMPK equation) \cite{Abrikosov1981, Mueller2011} fails completely. In particular, we find that the transmission distribution exhibits a bimodal structure typical of diffusive transport,  usually absent in one dimension. The characteristic universal features of Anderson localization re-emerge only for larger $\hbar_\text{eff}$, in particular in the deep quantum limit $1/\hbar_\text{eff}\ll1$ where usual, quantum-mechanical perturbative expansions at weak disorder can be applied \cite{Lugan2009}.

The article is organized as follows. Section \ref{Sec:model} recalls the main properties of the red-detuned random potential studied throughout this work. We define the semi-classical transport regime and derive the critical, classical scaling of the localization length at the percolation threshold. In Sec.~\ref{Sec:logtransmission}, we first attempt to connect Anderson localization theory to this classical limit by defining the localization length from the logarithm of individual speckle potential barriers. We show that this approach is inadequate and that accessing the classical regime requires a more appropriate definition. A semi-classical treatment valid for $\hbar_\text{eff}\ll1$ and resolving these limitations is then presented in Sec. \ref{Sec:SClocalization}, and benchmarked against exact transfer-matrix simulations. In Sec.~\ref{Sec:transmissionP}, we extend the analysis to the full statistical distribution of the particle transmission across the percolation threshold. In the subcritical region, this distribution becomes bimodal—a feature not captured by the standard 1D Fokker-Planck equation but well reproduced by our semi-classical approach. Finally, Sec.~\ref{Sec:SC_AL_crossover} discusses the crossover to the deep quantum regime ($\hbar_\text{eff}\gg1$), where perturbation theory in $1/\hbar_\text{eff}$ becomes applicable.

\section{Semi-classical transport in 1D red speckle potentials -- percolation transition}
\label{Sec:model}
\vspace{0.3cm}

\subsection{Speckle disorder and semi-classical regime}
\vspace{0.3cm}

\begin{figure}
 \centering       \includegraphics[width=0.95\textwidth]{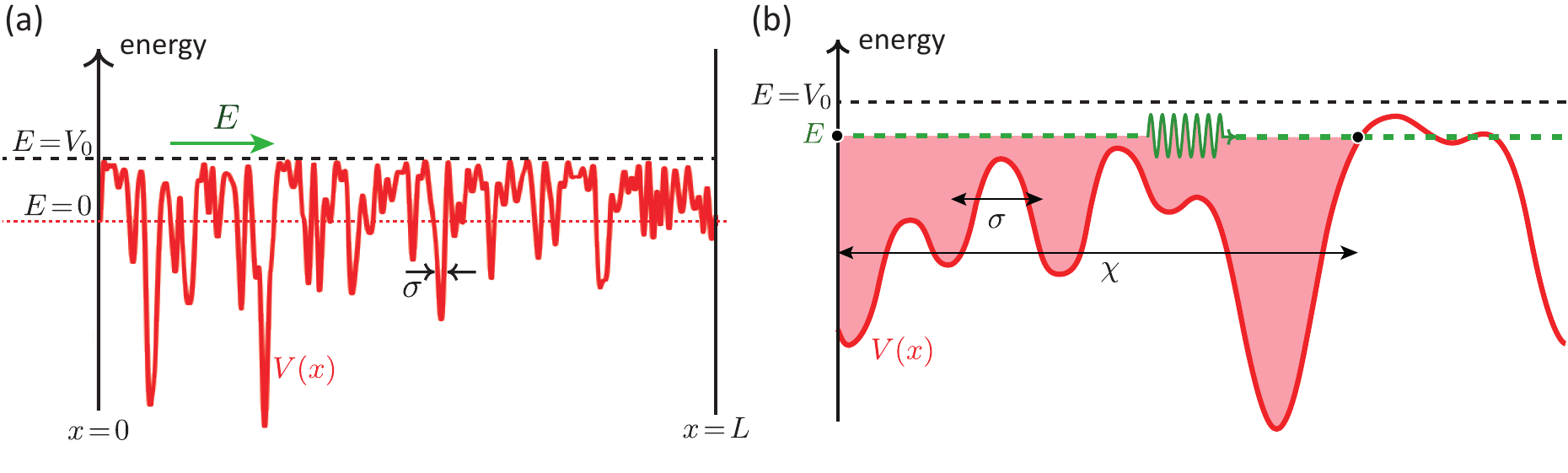}
 \caption{(a) The model: we consider the transmission of a quantum particle of energy $E\sim V_0\gg E_\sigma$ through a 1D~red-detuned speckle potential lying between $x=0$ and $x=L$. The potential $V(x)<V_0$ is bounded from above and has a correlation length $\sigma$. 
 (b) Near $E=V_0$, the speckle potential looks like a succession of inverted harmonic bumps of mean size $\sigma$. For a large enough system size, a classical particle with energy $E<V_0$ coming from the left always ends trapped within a potential valley (cluster). These clusters correspond to regions of space where $E-V(x)>0$, and have a characteristic size $\chi$. 
 }
\label{Fig:setup}
\end{figure}

We consider a quantum particle of mass $m$ and energy $E$ propagating through a region containing a ``red-detuned'' speckle random potential $V(x)$ between $x=0$ and $x=L$. This potential follows the Poissonian distribution \cite{Goodman2007}
\begin{equation}\label{Eq:Vdistribution}
P[V(x)]=\frac{1}{V_0}\exp\left[\frac{V(x)-V_0}{V_0}\right]\Theta(V_0-V),
\end{equation}
where $\Theta$ is the Heaviside function. This describes a potential bounded from above by $V_0>0$, with a mean  $\langle V(x)\rangle=0$ and a variance $\langle V^2(x)\rangle=V_0^2$. 
The speckle potential is further assumed to be Gaussian correlated:
\begin{equation}\label{Eq:Vcorrelation}
\langle V(x)V(x')\rangle=V_0^2\exp\left(-\frac{|x-x'|^2}{2\sigma^2}\right),
\end{equation}
where $\sigma$ denotes the correlation length. A typical realization of such a random potential is shown in Fig. \ref{Fig:setup}(a). In practice, 
correlated speckle potentials are commonly produced in cold-atom experiments by shining the atoms with a laser reflected or transmitted  through a rough surface. A red-detuned (resp. blue-detuned) speckle then corresponds to an attractive (resp. repulsive) potential experienced by the atoms, obtained by tuning the laser far to the red (resp. blue) of a relevant atomic transition~\cite{Clement2006, Shapiro2012}.

In a purely classical picture, it is clear that  the percolation threshold of this system lies at $E=V_0$. Indeed, for $E<V_0$ only finite clusters exist, whereas for $E>V_0$ an infinite cluster spans the entire space in the thermodynamic limit $L\to\infty$. From a transport perspective, a particle with energy $E<V_0$ is thus necessarily trapped within a potential valley, while for $E>V_0$ it can propagate arbitrarily far, see Fig. \ref{Fig:setup}(b). To explore how this classical percolation threshold connects with the  quantum theory of Anderson localization, in this work we focus---unless  otherwise stated---on the semi-classical regime defined by 
\begin{equation}
\label{eq:SC_condition}
\frac{\hbar^2}{m\sigma^2}\equiv E_\sigma\ll E\sim V_0.
\end{equation}
This regime corresponds to a physical situation where the particle flies near the classical percolation threshold, and has a de Broglie wavelength $h/\sqrt{2mE}$ much smaller than the disorder correlation length~$\sigma$. Under these conditions, the particle effectively resolves all potential fluctuations and experiences negligible tunneling, thereby behaving almost classically.
The condition~\eqref{eq:SC_condition} introduces  the relevant dimensionless parameter
\begin{equation}
\label{eq:hbareff_def}
\hbar_{\text{eff}}\equiv \sqrt{\frac{E_\sigma}{V_0}}\ll 1,
\end{equation}
which may be viewed as an effective Planck constant, quantifying how close the system is to the classical limit. In practice, this limit is typically reached either for very large disorder strength $V_0$, or for a sufficiently large correlation length $\sigma$.

\subsection{Percolation transition and critical scaling of transmission}
\label{Sec:classical_scaling}
\vspace{0.3cm}

In the strict classical limit $\hbar_\text{eff}=0$,  the percolation threshold at $E=V_0$ corresponds to a phase transition that manifests itself through the critical behavior of a characteristic cluster size. In this work, we are primarily interested in how this critical phenomenon appears in particle transport, which can be naturally characterized by the average  transmission coefficient $\ev{T}$ through the disordered region, where the brackets denote averaging over the speckle statistics. Near the percolation threshold, its classical value $\ev{T_\text{class}}$ is expected to take the form $\ev{T_\text{class}} \sim \exp[-{L}/{\chi(E)}]$, where $\chi(E)$ is a characteristic cluster size [see Fig. \ref{Fig:setup}(b)] behaving as 
\begin{equation}
\label{eq:chi_critical}
\chi(E\to V_0^-)\sim \frac{1}{(V_0-E)^\gamma}\ \ \text{and}\ \ \, \chi(E\geq V_0)=\infty,
\end{equation}
with $\gamma$ a critical exponent.  In terms of the topographic properties of the speckle potential, the factor $\exp(-{L}/{\chi})$ can alternatively be interpreted as the probability distribution of clusters of size~$L$: this distribution is essentially nonzero for $L<\chi(E)$, indicating that at a given energy $E$, most clusters have sizes smaller than $\chi(E)$.

To demonstrate Eq.~(\ref{eq:chi_critical}), we first note that for $E\simeq V_0$, the speckle potential can be viewed as a succession of inverted harmonic potential wells, characterized by random maxima $V$ located at positions $x_m$ satisfying $\partial_x V(x)|_{x=x_m}=0$,  and by random curvatures $\omega$ defined through $m\omega^2=-\partial^2_x V(x)|_{x=x_m}$. The detailed structure of the lower parts of the potential is of much lesser importance. For a speckle statistics, the joint distribution of potential maxima $V$ and curvatures $\omega$ was evaluated in Ref.~\cite{Prat2016}. For $0<V_0-V\ll V_0$, it takes the form
\begin{equation}
 \label{eq:PVomega}
P(V,\omega)\simeq\frac{2\sqrt{2}}{\omega_0}\frac{1}{\sqrt{\pi (V_0-V) V_0}}
\left(\frac{\omega}{\omega_0}\right)^2
\exp(-\frac{\omega^2}{\omega_0^2})\,,
\end{equation}
where $\omega_0\equiv \sqrt{V_0/(m\sigma^2)}$. This approximate expression will turn out to be sufficient for all subsequent calculations. We will also make use of the normalization condition:
\begin{equation}
\label{eq:normalization}
\int_0^\infty d\omega\int_{-\infty}^{V_0} dV\, P(V,\omega)=1 \,.
\end{equation}
Note that this normalization requires the full form of $P(V,\omega)$, available in \cite{Prat2016}, which unlike Eq.~\eqref{eq:PVomega} is integrable at $V\to-\infty$.
\newline

The distribution $P(V,\omega)$ can be used to compute $\ev{T_\text{class}}$, by noticing that near $E=V_0$ the classical transmission is the product of the average transmissions through uncorrelated inverted harmonic barriers, each of typical size $\sigma$. Denoting by $T_b$ the transmission coefficient through a single barrier of maximum $V$, we have $T_b=\Theta(E-V)$ since a classical particle is either transmitted or reflected depending on whether its energy $E$ is larger or smaller than $V$. The total transmission is therefore
\begin{equation}
\ev{T_\text{class}} = \ev{T_b}^{N_b} = \left[\int_0^\infty d\omega\int _{-\infty}^{V_0}dV P(V,\omega)\Theta(E-V)\right]^{N_b} \,.
\label{eq:Tclass_def}
\end{equation}
with $N_b$ the total number of barriers. Using Eq. (\ref{eq:normalization}), when 
$E<V_0$ this can be rewritten as:
\begin{equation}
\ev{T_\text{class}} = \left[1-\int_0^\infty d\omega\int _{E}^{V_0}dV P(V,\omega)\right]^{N_b}=\left[1-\sqrt{2(V_0-E)/V_0}\right]^{N_b} \,,
\end{equation}
where we have used Eq.~\eqref{eq:PVomega} to perform the last integral. Introducing the linear density of speckle maxima in the thermodynamic limit, $\rho \equiv N_b/L$, we finally obtain:
\begin{equation}
\label{eq:chiclassical}
\ev{T_{\text{class}}} = \exp[-\frac{L}{\chi(E)}],\ \ \ \chi(E)\underset{E\to V_0^-}{\simeq}\frac{1}{\rho \sqrt{2(V_0-E)/V_0}}.
\end{equation}
This is nothing but the critical scaling~\eqref{eq:chi_critical}, where the critical exponent $\gamma=1/2$. The density of maxima for a speckle potential was evaluated in Ref. \cite{Prat2016}, who found $\rho\simeq 0.2840/\sigma$.

\section{Localization length from logarithmic transmission}
\vspace{0.3cm}
\label{Sec:logtransmission}


Before investigating the fate of the percolation transition beyond the classical limit,  one may ask whether the classical result \eqref{eq:chiclassical} for $\chi$ can be recovered from the ``conventional'' theory of Anderson localization  in one dimension, describing propagation of a quantum particle through many barriers with complete phase randomization \cite{Abrikosov1981, Mueller2011}. Within this approach, which typically works for uncorrelated disorder and Gaussian statistics, the disorder-averaged transmission is predicted to scale with the system size as $\ev{T} \sim \exp[-L/4\xi(E)]$, with a localization length 
\begin{equation}
\label{eq:xidef}
\xi^{-1}(E)=-\rho\,\ev{\text{ln}\, T_b}
\end{equation}
defined from the logarithm of the transmission $T_b$ through a single potential barrier. A quick look at this formula highlights an obvious difficulty arising for energies $E<V_0$, if one tries to evaluate it in the classical limit $\hbar_\text{eff}=0$:
\begin{align}
\label{eq:xim1infty}
\xi^{-1}(E<V_0)
\underset{\hbar_\text{eff}=0}{\longrightarrow}
&-\rho\int_0^\infty  d\omega\int_{-\infty}^{V_0} dV\, P(V,\omega)\,\text{ln}\left[\Theta(E-V)\right]\nonumber\\
&=
-\rho\int_0^\infty  \!d\omega\!\int_{E}^{V_0} \!dV\, P(V,\omega)\text{ln}\left[\Theta(E-V)\right]
=
\infty.
\end{align}
In other words, Eq.~\eqref{eq:xidef} predicts a strict vanishing of the localization length in the classical regime, which is clearly incompatible with Eq.~\eqref{eq:chiclassical}.

\begin{figure}[h]
 \centering
 \includegraphics[width=0.55\textwidth]{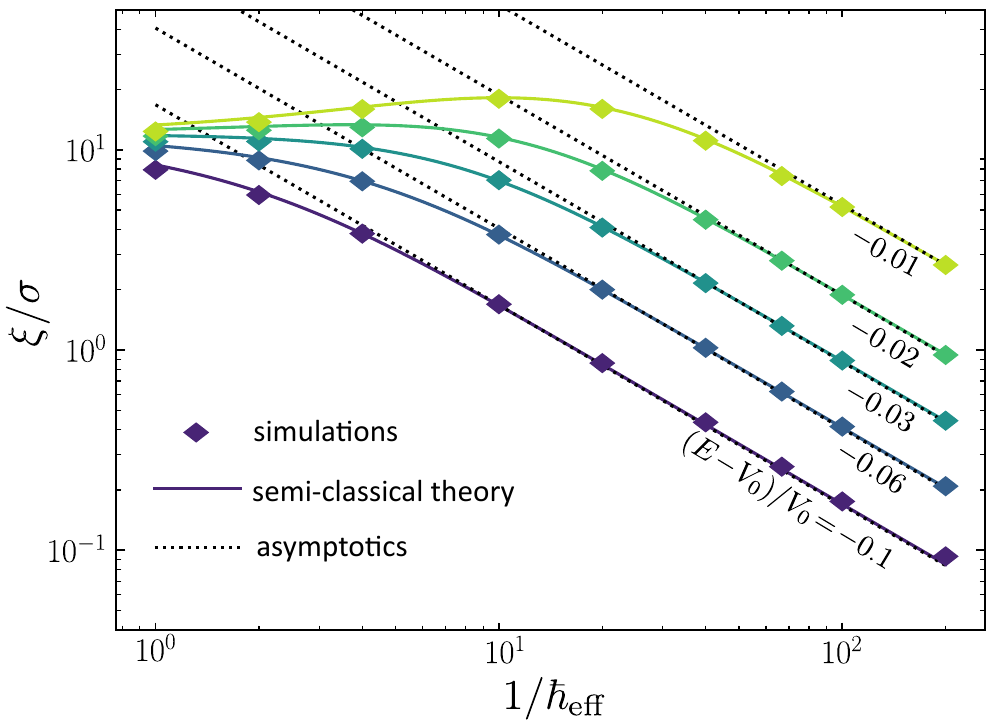}
 \caption{Localization length $\xi \equiv -L / \ev{\ln T}$, numerically computed from the average logarithmic transmission using the transfer-matrix method, as a function of $1/\hbar_\text{eff}$ for several energies below the percolation threshold (discrete points). The plot shows that $\xi \to 0$ as $\hbar_\text{eff} \to 0$, i.e., does not converge to the classical result~\eqref{eq:chiclassical}. The solid curves represent the semi-classical prediction (\ref{eq:xim1_SC}), and the dotted lines  the asymptotic formula~\eqref{Eq:SC_asymptotics}. In the transfer-matrix simulations, $\xi$ is obtained from linear regressions of $\ev{\ln T}$ computed for 20 system sizes between $L=5 \sigma$ and $L=100\sigma$, each averaged over $10^4$ disorder realizations.
  }
\label{Fig.xiE}
\end{figure}

To gain deeper insight into the problem, it is also instructive to examine the behavior of $\xi(E<V_0)$ for $\hbar_\text{eff}$ small but finite. Figure~\ref{Fig.xiE} presents numerical results in this regime, obtained using the transfer-matrix method in a 1D~red-detuned speckle potential (see Sec.~\ref{Sec:SClocalization} for details on this approach). The data show that $\xi$ remains finite for finite values of $\hbar_\text{eff}$, but indeed vanishes—rather than approaching a finite limit—as $\hbar_\text{eff} \to 0$.
This observation can be directly confirmed by an analytical calculation of~\eqref{eq:xidef}, using the quantum-mechanical expression of the transmission $T_b$ through an inverted harmonic potential barrier with maximum $V$ and curvature $\omega$. For such a barrier the calculation can be done exactly~\cite{Holstein1982, Maitra1996}:
\begin{equation}
\label{eq:Ti_IHO}
T_b=\frac{1}{1+\exp[-\frac{2\pi}{\hbar\omega}(E-V)]} \,.
\end{equation}
Inserting this expression into Eq.~\eqref{eq:xidef}, we get:
\begin{equation}
\label{eq:xim1_SC}
\xi^{-1}(E)=\rho \int_{0}^\infty d\omega\int_{-\infty}^{V_0} dV\, P(V,\omega)\,
\ln(1+\exp[-\frac{2\pi}{\hbar\omega}(E-V)]).
\end{equation}
Let us focus on energies $E<V_0$ below the percolation threshold. The integral in~\eqref{eq:xim1_SC} can be evaluated by separating the contributions of clusters ($E>V$) and tunneling ($E<V$), and further using that $\hbar_\text{eff}\ll 1$. This leads to
\begin{equation}
\xi^{-1}(E)\simeq \rho \int_{0}^\infty d\omega\left\{
\int_{-\infty}^{E} dV\, \exp[-\frac{2\pi}{\hbar\omega}(E-V)]
+ \int_{E}^{V_0} dV\frac{2\pi}{\hbar\omega}(V-E)
\right\}P(V,\omega)\,.
\end{equation}
In this expression, the contribution of clusters (first term on the right-hand side) is completely negligible compared to the contribution of tunneling (second term) when $\hbar_\text{eff}\propto \hbar\ll 1$. Making use of Eq.~\eqref{eq:PVomega}, we then find:
\begin{equation}
\label{Eq:SC_asymptotics}
\xi^{-1}(E)\simeq\frac{8\rho \sqrt{2\pi}}{3\hbar_\text{eff}} \left(\frac{V_0-E}{V_0}\right)^{3/2}\underset{\hbar_\text{eff}\to0}{\longrightarrow} \infty.
\end{equation}
This relation, together with the exact formula~\eqref{eq:xim1_SC}, is shown in Fig.~\ref{Fig.xiE} and matches well the transfer-matrix computation of~$\xi$ defined from the average logarithmic transmission.\newline

Both the classical result~\eqref{eq:xim1infty} or its semi-classical version~\eqref{Eq:SC_asymptotics} thus confirm that the localization length defined according to~\eqref{eq:xidef} does \emph{not} converge to $\chi$ [Eq.~\eqref{eq:chiclassical}], which is the proper critical behavior  expected at the percolation threshold. The reason for this mismatch is evident from the above calculation: the localization length defined via~\eqref{eq:xidef} describes a propagation dominated by tunneling through barriers, while completely neglecting the cluster contributions $E>V$. This is legitimate in the deep quantum regime $\hbar_{\text{eff}}\gg 1$ (see Sec.~\ref{Sec:deep_quantum}), but is clearly inadequate in the semi-classical regime~\eqref{eq:SC_condition} where the particle energy, on the contrary, exceeds the height of most encountered barriers. From a mathematical standpoint, the reason why Eq.~\eqref{eq:xidef} incorrectly captures the decay of $\langle T\rangle$ when $\hbar_\text{eff}\ll1$ lies in the fact that it originates from a model in which disorder is assumed to be both \emph{uncorrelated} and \emph{Gaussian} distributed~\cite{Mueller2011, Abrikosov1981}. These assumptions are fundamentally incompatible with the semi-classical condition~\eqref{eq:SC_condition} for two reasons. First, because $\hbar_\text{eff}$ diverges as the disorder correlation length $\sigma\to0$, making the classical condition $E\gg E_\sigma$ never fulfilled. Second, because near the upper bound $E\sim V_0$, the non-Gaussian character of the red-detuned speckle potential becomes most pronounced, while a Gaussian model would instead describe a potential with arbitrarily high peaks, obviously not bounded. In the next section, we present a correct determination of~$\ev{T}$ and of the localization length in the semi-classical limit.

\section{Semi-classical localization length}
\vspace{0.3cm}
\label{Sec:SClocalization}

\subsection{Transfer-matrix simulations}\label{sec:transfermatrix}
\vspace{0.3cm}

To evaluate the mean transmission $\ev{T}$ in the semi-classical regime, we first perform numerical simulations based on the 1D~transfer-matrix method. We discretize the stationary Schr\"odinger equation $-\hbar^2\partial_x^2\psi(x)/(2m)+V(x)\psi(x)=E\psi(x)$ with a red-detuned potential $V(x)$ over a system of length~$L$. Expressing the wave function $\psi = \{\psi_n\}_{0 \leq n < N}$ and the potential $V = {V_n}$ on $N = L/a$ lattice sites, we obtain the effective tight-binding model:
\begin{equation}\label{Eq:Schrödinger discrete}
    J  \psi_{n+1}= \left(V_n-E-2J \right)\psi_n-J\psi_{n-1}\,,
\end{equation}
where $J = \hbar^2/(2ma^2)$. To ensure that the simulations accurately reproduce propagation in a continuous random potential, the lattice spacing $a$ must be chosen much smaller than both the disorder correlation length $\sigma$ and the de Broglie wavelength. 
In the semi-classical regime, the
latter scales as $h/\sqrt{2mE} \sim h/\sqrt{2mV_0} \sim \hbar_\text{eff}\sigma$.  We therefore choose $a = \hbar_\text{eff}\sigma/4$, which is typically sufficient for accurate calculations up to $\hbar_\text{eff} \sim 1$. To numerically generate a potential with  distribution~\eqref{Eq:Vdistribution} and correlation function~\eqref{Eq:Vcorrelation}, we convolve an uncorrelated, complex Gaussian random field  $\mathcal E(x)$ with the kernel $c(x)=V_0 \exp({-x^2/4\sigma^2})$. The red-detuned speckle potential is then obtained as $V(x)=-|\mathcal E(x)|^2$. To compute the transmission through the chain, we propagate Eq.~\eqref{Eq:Schrödinger discrete} backward for a given disorder realization and system size $L$, starting from the right boundary condition $\psi_{N-1}=1$ and $\psi_{N}=e^{ika}$. This choice describes an outgoing plane wave of momentum $k>0$ in a disorder-free region, with energy given by the tight-binding dispersion relation $E = 2J(1-\cos ka)$. The transmission coefficient $T$ follows from the standard transfer-matrix formalism as the ratio of outgoing to incoming plane-wave amplitudes:
\begin{equation}\label{Eq:Transmission}
    T = \left| \frac{2\sin ka}{\psi_0 - e^{-ika}\,\psi_{-1}} \right|^2,
\end{equation}
which we eventually average over many (typically $10^4$) realizations of the random potential.\newline

\begin{figure}[h]
 \centering
 \includegraphics[width=1\textwidth]{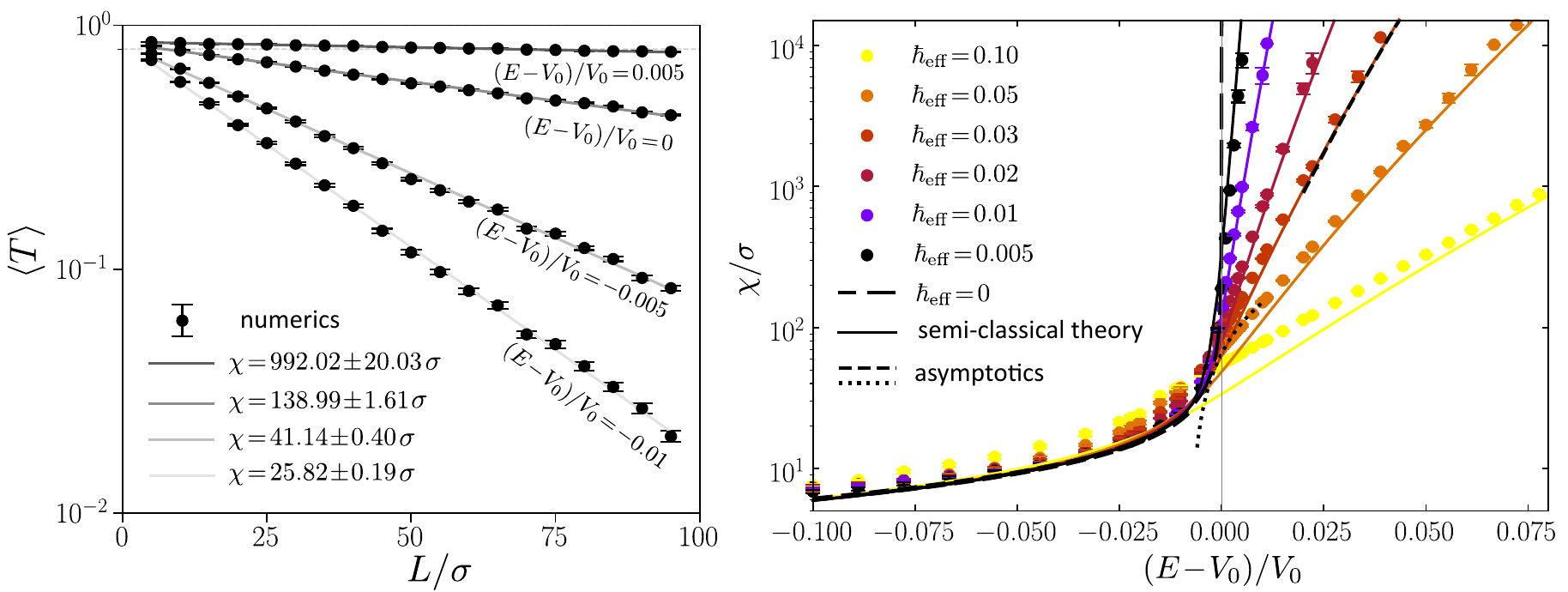}
 \caption{
 \label{Fig:chi_E}
 (a) Average transmission $\ev{T}$ as a function of system size, computed numerically from the transfer-matrix method (black points) for $\hbar_\text{eff}=0.01$, for  energies near the classical percolation threshold. Solid lines are fits to $\exp(-L/\chi)$. (b) Localization length $\chi(E)$ as a function of energy for several values of~$\hbar_\text{eff}$, deduced from  the transfer-matrix simulations (colored points). The solid curves are the theoretical prediction~\eqref{eq:chitheory} at finite $\hbar_{\text{eff}}$. The long-dashed curve is the classical limit~\eqref{eq:chiclassical}, and the dashed and dotted curves the asymptotic formulas~\eqref{eq:chi_asymptote1} and \eqref{eq:chi_asymptotics2}, respectively, shown for $\hbar_\text{eff}=0.03$. Here $\chi$ is obtained from linear regressions of $\ev{T}$ computed for 20~system sizes between $L=5 \sigma$ and $L=100\sigma$, each averaged over $10^4$ disorder realizations.
 }
\label{Fig:chiE}
\end{figure}
Figure~\ref{Fig:chiE}(a) first shows the average transmission as a function of the system size~$L$, computed numerically for $\hbar_\text{eff}=0.01$ and for energies in the vicinity of the percolation threshold. The transmission exhibits a marked exponential decay. Fits of the data to $\propto \exp[-L/\chi(E)]$ allow us to extract the localization length~$\chi$ as a function of energy, which is plotted in Fig. \ref{Fig:chi_E}(b) for several values of $\hbar_\text{eff}$ closer and closer to the classical limit. Remarkably, as $\hbar_\text{eff}\to 0$, the localization length converges to the classical result~\eqref{eq:chiclassical} (long-dashed curve) that characterizes the percolation transition. The figure also reveals how this classical phase transition disappears for finite~$\hbar_\text{eff}$: the function~$\chi(E)$ smoothly evolves from a strict algebraic divergence at $E=V_0$ when $\hbar_\text{eff}=0$, to an ultra-steep yet continuous behavior at this point as soon as $\hbar_\text{eff}$ is nonzero. 
This is the first important result of the paper.

\subsection{Semi-classical theory}
\vspace{0.3cm}

To explain the numerical results of Fig.~\ref{Fig:chiE}(b), we now present a theoretical calculation of the localization length in the semi-classical regime. To this end, we first note that in the deep quantum regime $\hbar_\text{eff}\gg1$ of Anderson localization, statistical correlations between transmission events typically arise from the accumulation of phase factors associated with multiple quantum-mechanical reflections between successive barriers. For energies $E\simeq V_0$ in the semi-classical regime $\hbar_\text{eff}\ll1$, in contrast, such reflections are negligible because the transmission through each barrier is very close to unity. As a result, transmission events $T_b$ through individual barriers are essentially independent, and
\begin{equation}
\label{eq:TapproxSC}
    \ev{T} \simeq \ev{T_b}^{N_b}
\end{equation}
like in the classical limit [see Eq.~\eqref{eq:Tclass_def}]. 
Under this approximation, the average transmission in the semi-classical regime is given by
\begin{equation}
\label{eq:TSC_def}
    \ev{T} \simeq \ev{T_b}^{N_b} \simeq \exp[-L/\chi(E)],\ \ \chi^{-1}(E)\equiv -\rho\,\ln \ev{T_b},
\end{equation}
where we have used that the total number of barriers $N_b=\rho L$, with $\rho$ the density of speckle maxima. As compared to Eq.~\eqref{eq:xidef}, observe that in this relation the logarithm is taken \emph{after} the disorder average is performed. Using Eq.~\eqref{eq:Ti_IHO}, we obtain the explicit expression of the semi-classical localization length:
\begin{equation}
\chi^{-1}(E)=-\rho\,\text{ln} \int_{0}^\infty d\omega\int_{-\infty}^{V_0} dV\, \frac{P(V,\omega)}{
1+\exp\left[-\frac{2\pi}{\hbar\omega}(E-V)\right]}.
\end{equation}
To simplify this formula, we use that $1/[1+\exp(x)]=1-1/[1+\exp(-x)]$, together with the normalization condition~\eqref{eq:normalization}:
\begin{equation}\label{eq:chitheory_valid_for_any_E}
\chi^{-1}(E)=-\rho\,\text{ln} 
\left\{1-
\int_{0}^\infty d\omega\int_{-\infty}^{V_0} dV\, \frac{P(V,\omega)}{1+\exp[\frac{2\pi}{\hbar\omega}(E-V)]}
\right\}.
\end{equation}
An advantage of this relation is that the integral over $V$ is now convergent with the approximate expression~\eqref{eq:PVomega} for $P(V,\omega)$. Performing this integral, we obtain
\begin{equation}
\label{eq:chitheory}
\chi^{-1}(E)=-\rho\,\text{ln} 
\left\{1+\frac{2\sqrt{\hbar_\text{eff}}}{\sqrt{\pi}}
\int_{0}^\infty d\tilde\omega\, 
\tilde\omega^{5/2}e^{-\tilde\omega^2}
\text{Li}_{1/2}
\left(-\exp[\frac{2\pi (V_0-E)}{\hbar_\text{eff}V_0\tilde\omega}]\right)
\right\}\,,
\end{equation}
where $\text{Li}_{1/2}(x)$ is the polylogarithm of order 1/2, and we have introduced the dimensionless variable $\tilde\omega\equiv \omega/\omega_0$ and the effective Planck constant (\ref{eq:hbareff_def}). In Fig. \ref{Fig:chiE}(b), we compare the localization length computed with Eq.~\eqref{eq:chitheory} to the exact transfer-matrix simulations for different values of $\hbar_\text{eff}\ll1$. The agreement with the numerics is excellent up to $\hbar_\text{eff}\simeq 0.05$, demonstrating that Eqs.~\eqref{eq:TSC_def} and~\eqref{eq:chitheory} provide the correct description of the localization length in the semi-classical regime. 

\subsection{Asymptotic regimes}
\vspace{0.3cm}

To gain further insight into the behavior of $\chi(E)$, it is instructive to derive from Eq.~\eqref{eq:chitheory} approximate  expressions above, at, and below the percolation threshold.

\subsubsection{Localization length above the threshold}

We first consider the regime of energies $E$ well above the percolation threshold $V_0$. In this case, the factor within the exponential in Eq.~\eqref{eq:chitheory} typically takes large negative values. Using that $\text{Li}_{1/2}(-x)\simeq -x$ when $x\to0$, we infer:
\begin{equation}
    \chi^{-1}(E>V_0)\simeq
    -\rho\ln\left\{
    1-\frac{2\sqrt{\hbar_\text{eff}}}{\sqrt{\pi}}
    \int_0^{\infty}d\tilde\omega \,\tilde\omega^{5/2}
    \exp[-\tilde\omega^2 +\frac{2\pi (V_0-E)}{\hbar_\text{eff}V_0\tilde\omega}]
    \right\}.
    \label{eq:chi_asymptote1}
\end{equation}
The remaining integral can be evaluated using the saddle-point approximation. This yields the asymptotic relation
\begin{equation}
\label{eq:saddle_point}
\chi(E>V_0)\simeq\frac{\sqrt{3}\,\hbar_\text{eff}^{1/3}}{2\rho [(E-V_0)\pi/V_0]^{5/6}}\exp\left\{\,3\left[\frac{(E-V_0)\pi}{\hbar_\text{eff}V_0}\right]^{2/3}\right\}.
\end{equation}
Above the percolation threshold, the localization length thus grows as a stretched exponential of the energy and exhibits a non-analytic dependence on $\hbar_\text{eff}$, in agreement with the numerical observations. The expression~\eqref{eq:chi_asymptote1}—which is slightly more accurate than Eq.~\eqref{eq:saddle_point}—is shown in Fig.~\ref{Fig:chiE}(b) together with the exact numerical data (dashed curve), and a very good agreement is found.

\subsubsection{Localization length near the threshold}

Near the percolation threshold $E\simeq V_0$, Eq.~\eqref{eq:chitheory} can be simplified by using the Taylor expansion $\text{Li}_{1/2}(-x)\simeq (\sqrt{2}-1)\zeta(1/2)+(2\sqrt{2}-1)\zeta(-1/2)(x-1)$ of the polylogarithm when $x\to1$. Inserting this into Eq.~\eqref{eq:chitheory} and performing the integrals over $\tilde\omega$, we find
\begin{equation}
\label{eq:chi_asymptotics2}
    \chi(E\simeq V_0)\simeq\frac{1}{\rho\alpha\sqrt{\hbar_\text{eff}}}+\frac{\beta}{\rho \alpha^2}\frac{E-V_0}{V_0\hbar_\text{eff}^{3/2}}+\mathcal{O}((E-V_0)^2),
\end{equation}
where $\alpha=(1-\sqrt{2})\zeta(1/2)\Gamma(7/4)/\sqrt{\pi}$ and $\beta=2\sqrt{\pi}(1-2\sqrt{2})\zeta(-1/2)\Gamma(5/4)$. The localization length thus diverges as $\smash{\hbar_\text{eff}^{-1/2}}$ at the percolation threshold. The approximation~\eqref{eq:chi_asymptotics2} is shown as a dotted curve in Fig.~\ref{Fig:chiE}(b).

\subsubsection{Localization length below the threshold} 

We finally turn our attention to the most interesting regime: the behavior of localization below the threshold. To obtain an approximate expression for~$\chi$ in this case, we use the asymptotic expansion $\text{Li}_{1/2}(-x)\simeq -2\sqrt{\ln(x)/\pi}+\pi^{3/2}/(12\,\ln^{3/2}x)$ when $x\to\infty$ in Eq.~\eqref{eq:chitheory}. In terms of the inverse localization length, this leads to:
\begin{equation}
\label{eq:chi_asymptotics3}
    \chi^{-1}(E<V_0)\simeq 
    \rho\sqrt{2(V_0-E)/V_0}
    -\frac{\rho\,\hbar_\text{eff}^2}{16[2(V_0-E)/V_0]^{3/2}}+\mathcal{O}((V_0-E)^{-7/2}).
\end{equation}
At zeroth order in $\hbar_\text{eff}$, this result reduces to the classical expression~\eqref{eq:chiclassical}, as expected. The next term, proportional to $\hbar_\text{eff}^2$, represents the first semi-classical correction at energies well below the percolation threshold.\newline

Together with the exact relation (\ref{eq:chitheory}), Eqs.~\eqref{eq:saddle_point}, \eqref{eq:chi_asymptotics2} and \eqref{eq:chi_asymptotics3} provide a complete description of the localization length near the percolation threshold in the semi-classical limit. In the next section, we extend this analysis to another important observable, the statistical distribution of transmission.

\section{Statistics of transmission}
\vspace{0.3cm}
\label{Sec:transmissionP}

\subsection{Classical limit}
\vspace{0.3cm}

To further characterize transport across the percolation transition in the semi-classical regime, let us also examine the statistical distribution of the transmission, $P(T)$, for small values of $\hbar_\text{eff}$. In the classical limit $\hbar_\text{eff}=0$, a straightforward calculation presented in the appendix yields
\begin{equation}
\label{eq:ptclassical}
    P(T)=\delta(T-1)
\end{equation}
for energies $E\geq V_0$, i.e., at and above the percolation threshold, and
\begin{equation}
\label{eq:doubledelta}
    P(T)=\ev{T_\text{class}} \, \delta(T-1) + (1-\ev{T_\text{class}})\,\delta(T)
\end{equation}
in the subcritical regime $E\leq V_0$, where the average classical transmission $\ev{T_\text{class}}$ is given in Eq.~\eqref{eq:chiclassical}. These expressions directly imply that $\smash{\ev{T} \equiv \int_0^1 T P(T)dT=1}$ for $E\geq V_0$, and $\smash{\ev{T}=\ev{T_\text{class}}}$ for $E\leq V_0$. In the classical limit, the percolation transition therefore appears as a clear discontinuity of $P(T)$ at $E=V_0$, with an intuitive physical interpretation: Above the transition, the particle is fully transmitted regardless of the disorder realization. Below the transition, for any given realization, the particle is either entirely transmitted or completely reflected. The coefficients $\ev{T_\text{class}}$ and $1-\ev{T_\text{class}}$ can thus be interpreted as the fractions of realizations leading to full transmission and full reflection, respectively.

\subsection{Numerical simulations at finite $\hbar_\text{eff}$}
\vspace{0.3cm}

\begin{figure}[h]
 \centering
 \includegraphics[width=0.55\textwidth]{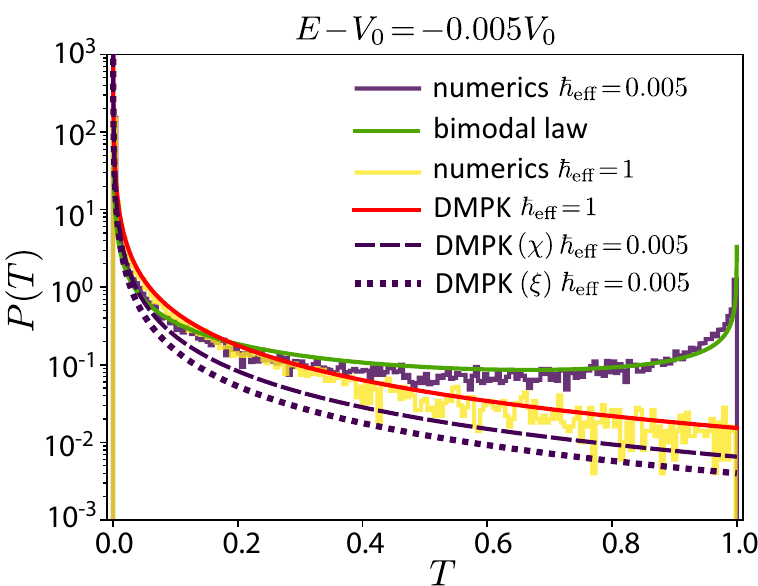}
 \caption{Transmission distribution $P(T)$ below the percolation threshold ($E-V_0=-0.005V_0$), in the semi-classical regime with $\hbar_\text{eff}=0.005$. A comparison between the numerical results and the DMPK prediction~\eqref{eq:abrikosov}, using either $\xi$ or $\chi/4$ as the localization length (dotted and dashed curves), highlights the inadequacy of this approach. In contrast, the comparison with the bimodal law~\eqref{eq:bimodal} (solid green curve) shows excellent agreement. For reference, we also display $P(T)$ in the deep quantum regime ($\hbar_\text{eff}=1$), for which the DMPK equation becomes accurate again (solid red curve). All simulations are performed for a system size $L=100\sigma$ and involve binning over $10^5$ disorder realizations.
  }
\label{Fig.PTbelow}
\end{figure}

We now turn to the transmission distribution at finite $\hbar_\text{eff}$. Like for the localization length, we expect a smoothing of $P(T)$ across the percolation threshold. Transfer-matrix simulations of $P(T)$ shown in  Fig. \ref{Fig.PT3} for three energies around $E=V_0$ confirm this statement:
\begin{figure}[h]
 \centering
 \includegraphics[width=0.55\textwidth]{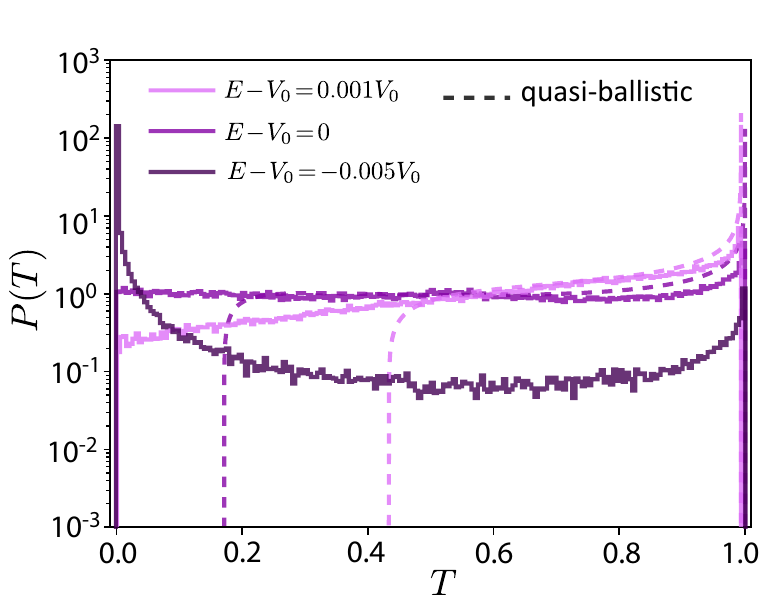}
 \caption{Transmission distribution $P(T)$ below, at and above the percolation threshold, computed numerically in the semi-classical regime with $\hbar_\text{eff}=0.005$. The simulations are performed for a system size $L=100\sigma$ and involve binning over $10^5$ disorder realizations. The dashed curves show the quasi-ballistic formula~\eqref{eq:qb} in the two cases $E=0$ and $E> V_0$. 
  }
\label{Fig.PT3}
\end{figure}
As the threshold is crossed for finite $\hbar_\text{eff}$, $P(T)$ evolves continuously from a bimodal form for $E<V_0$ toward a distribution sharply peaked at $T=1$ for $E\geq V_0$. The most intriguing phenomenon here occurs below the threshold, where the classical doubled-delta law~\eqref{eq:doubledelta} is replaced, in the semi-classical regime, by a much smoother bimodal distribution. Remarkably, this bimodal structure is \emph{not} captured by conventional 1D localization theory for uncorrelated Gaussian disorder. To illustrate this failure, we reproduce in Fig.~\ref{Fig.PTbelow} the numerically obtained bimodal distribution for $E< E_0$, and compare it with the exact solution of Abrikosov's Fokker-Planck (or 1D DMPK) equation for the transmission distribution derived for Gaussian and uncorrelated disorder~\cite{Abrikosov1981},
\begin{equation}
\label{eq:abrikosov}
    P(T)=\frac{2}{T^2\sqrt{\pi}(L/\xi)^{3/2}} \int_{\text{acosh}(1/\sqrt{T})}^\infty
    \frac{x\exp(-x^2\xi/L-L/4\xi)}{\sqrt{\cosh^2x-1/T}} \, {\rm d} x.
\end{equation}
\newline
This prediction, shown as a dotted line in Fig.~\ref{Fig.PTbelow}, is clearly incompatible with our  numerical results. Even replacing $\xi$  by the correct localization scale $\chi/4$ fails to reproduce the simulations.
For reference, we also show in Fig.~\ref{Fig.PTbelow} a numerical calculation of~$P(T)$ in the quantum regime $\hbar_\text{eff}=1$. In this case, $P(T)$ takes the more usual log-normal shape and Eq.~\eqref{eq:abrikosov} becomes accurate again. Like for the localization length, we attribute the breakdown of Eq.~\eqref{eq:abrikosov} in the semi-classical regime to the strong influence of spatial correlations and non-Gaussian statistics of the red-detuned disorder near the percolation threshold. In the quantum regime, tunneling through barriers dominates, making the disorder correlations less relevant and restoring the applicability of the DMPK theory developed for uncorrelated Gaussian potentials.

Turning back to the semi-classical regime, the shape of $P(T)$ strongly resembles the bimodal distribution characteristic of diffusive transport in multi-channel waveguides \cite{Beenakker1997}. It is important to stress, however, that the bimodality here has a fundamentally different origin a priori. In quasi-1D waveguides, diffusion indeed stems from the presence of many transverse channels $N$; the localization length  $\xi\sim N\ell\gg \ell$ is much longer than the mean free path $\ell$, so a diffusive window appears for system sizes $\ell\ll L\ll \xi$. In the present work, however, the random potential is strictly 1D ($N=1$)  so diffusion cannot be attributed to a multi-channel effect. Understanding how this bimodal distribution emerges from a semi-classical description and what is the diffusive mechanism behind it is the objective of the next section.

\subsection{The bimodal law from semi-classical theory}
\vspace{0.3cm}

We now show how the semi-classical theory can be used to derive the bimodal transmission distribution observed near the percolation threshold in Fig.~\ref{Fig.PTbelow}. As before, we express the total transmission $T\simeq \Pi_{j=1}^{N_b} T_j$ through the speckle in terms of the individual barrier transmissions $T_j$. Since dealing directly with the distribution of a product of random variables is inconvenient, we instead consider its logarithm, $\ln T=\ln T_1+\ldots \ln T_{N_b}$. The characteristic function of a sum of independent random variables being equal to the product of the individual characteristic functions~\cite{Dettmann2009}, we obtain
\begin{equation}
\label{eq:PlnT}
    P(\ln T)=\int_{-\infty}^\infty dt\,
    e^{-it\ln T}[\,\varphi(t)]^{N_b},
\end{equation}
where $\varphi(t)$ denotes the characteristic function of the random variable $\ln T_j$. Transforming back to~$T$, Eq.~\eqref{eq:PlnT} becomes:
\begin{equation}
\label{Eq:PT2}
    P(T)=\frac{1}{T}\int_{-\infty}^\infty {\rm d}t \, \exp\big[\!-it\ln T+N_b\ln \varphi(t) \big],
\end{equation}
which we evaluate using the saddle-point approximation. The saddle-point $t=t_\text{sp}$ is defined by
\begin{equation}
\label{eq:SP}
    \frac{\partial_t \varphi(t)}{\varphi(t)}\Big|_{t=t_\text{sp}} = \frac{i\ln T }{N_b} \,,
\end{equation}
and allows us to rewrite Eq.~\eqref{Eq:PT2} as 
\begin{equation}
\label{eq:PTSP}
    P(T)\sim\frac{1}{T} \frac{\exp\big[N_b\, \ln\varphi(t)-it\ln T\big]}
    {\sqrt{|N_b \partial_t^2 \ln\varphi(t)|}}\Big|_{t=t_\text{sp}}.
\end{equation}
To solve Eqs.~\eqref{eq:SP}–\eqref{eq:PTSP}, we need to evaluate the characteristic function
\begin{equation}
    \varphi(t)\equiv
    \ev{ e^{it\ln T_j} } = \int_0^\infty d\omega\int_{-\infty}^{V_0} dV\, P(V,\omega)
    \exp[ \!-it\ln( 1+e^{-\frac{2\pi}{\hbar\omega}(E-V)} ) ],
\end{equation}
where the argument of the exponential follows from Eq.~\eqref{eq:Ti_IHO} for the transmission through a single barrier. We now focus on the subcritical regime $E<V_0$. For small $\hbar$, the integral over barrier maxima~$V$ is dominated by the cluster contributions $E>V$. This leads to
\begin{equation}
    \varphi(t)\simeq
    \int_0^\infty d\omega\int_{-\infty}^{E} dV\, P(V,\omega)
    \exp[ \!-it e^{-\frac{2\pi}{\hbar\omega}(E-V)} ],
\end{equation}
where we further expanded the logarithm for small $\hbar$. Expressing the exponential factor as a power series and separating out the zeroth-order contribution yields
\begin{equation}
    \varphi(t)\simeq
    \int_0^\infty d\omega
    \sum_{n=1}^\infty
    \frac{(-it)^n}{n!}
    \int_{-\infty}^{E} dV\, P(V,\omega)
    e^{-\frac{2\pi n}{\hbar\omega}(E-V)}+
    \langle T_b\rangle,
\end{equation}
where we used that $\int_0^\infty d\omega\int_{-\infty}^{E} dV\, P(V,\omega)=1-\sqrt{2(V_0-E)/V_0}\equiv \ev{T_b}$. The first term on the right-hand side is finally evaluated in the limit of small $\hbar$ using Eq.~\eqref{eq:PVomega}. This leads to
\begin{equation}
\label{eq:solphi}
    \varphi(t)\simeq
    \ev{T_b} -\frac{\hbar_\text{eff}}{\pi^{3/2}\sqrt{2(V_0-E)/V_0}}
    \big[\gamma+\Gamma(0,it)+\ln(it)\big],
\end{equation}
where $\gamma$ is the Euler–Mascheroni constant and $\Gamma(0,x)$ the upper incomplete gamma function. With this result, the saddle-point equation~\eqref{eq:SP} becomes, to leading order in $\hbar_\text{eff}$,
\begin{equation}
    \frac{\hbar_\text{eff}}{\pi^{3/2}\sqrt{2(V_0-E)/V_0}}
   \frac{\exp(-it_\text{sp})-1}{t_\text{sp}}=\frac{i\ln T}{N_b} \langle T_b\rangle.
\end{equation}
For small $\hbar_\text{eff}$, the saddle-point solution of this equation is of the form $t_\text{sp}=i|t_\text{sp}|$, where $|t_\text{sp}|$ satisfies $e^{|t_\text{sp}|}/|t_\text{sp}|\simeq -\ev{T_b} \sqrt{2(V_0-E)/V_0}\pi^{3/2}\ln T/(N_b\hbar_\text{eff})$. Using this solution together with Eq.~\eqref{eq:solphi}, we infer:
\begin{align}
    \varphi(t_\text{sp})\simeq \langle T_b\rangle+\mathcal{O}(\hbar_\text{eff}), \ \ \   \partial_t^2\varphi(t)|_{t=t_\text{sp}}\simeq\frac{\ln T}{N_b}\langle T_b\rangle+\mathcal{O}(\hbar_\text{eff}).
\end{align}
Inserting these expressions into Eq.~\eqref{eq:PTSP}, and using that $\ev{T_b}^{N_b}\equiv \ev{T}$ and $-\ln T\simeq 1-T$ near $T=1$, we  finally obtain (up to a prefactor enforced by the condition $\smash{\int_0^1 dT\, TP(T)=\ev{T}}$):
\begin{equation}
\label{eq:bimodal}
    P(T)\simeq\frac{\langle T\rangle}{2T\sqrt{1-T}}.
\end{equation}
This result exactly coincides with the  bimodal law characteristic of diffusive systems~\cite{Beenakker1997}, which we here recover  from a semi-classical approach in the vicinity of the percolation threshold. This is the second important result of the paper. Equation~\eqref{eq:bimodal} is shown in Fig.~\ref{Fig.PTbelow}, and describes very well the transfer-matrix simulations, even up to rather small transmission values.  

Physically, we conjecture that the bimodal distribution~\eqref{eq:bimodal} is associated with a diffusive process in which the particle of energy~$E$ is multiply scattered by barriers whose height is close to $E$. For such barriers, the transmission coefficient is broadly distributed around 1/2 due to quantum reflection, while the peaks at $T=0$ and $T=1$ are associated with events in which the particle is either purely reflected or purely transmitted. Within this picture, the characteristic length~$\chi $ plays a role analogous to a mean free path.

\subsection{Quasi-ballistic regime}
\vspace{0.3cm}

For energies $E\geq V_0$, the numerical results in Fig.~\ref{Fig.PT3} show that the transmission distribution consists of a sharp peak near $T=1$ and a much smaller but long tail extending to lower $T$ values. While this structure might in principle be accessed from the semi-classical theory as well, we were not able to derive a simple result for this case. In Fig.~\ref{Fig.PT3}, we nevertheless show for comparison the ``quasi-ballistic'' result derived in~\cite{Beenakker1997} for multi-channel waveguides:
\begin{equation}
\label{eq:qb}
    P(x,t)=\frac{2}{\pi}\text{Im}\, U(x,t),\quad U(\zeta,t)=\text{coth}[\zeta-t\, U(\zeta,t)],
\end{equation}
 where $T=1/\cosh^2x$ and taking $t=L/\chi$. This relation leads to a reasonable agreement of the numerical results, at least for not too small $T$ values.

\section{From semi-classical to deep quantum regime}
\vspace{0.3cm}
\label{Sec:SC_AL_crossover}

So far, we have focused on the semi-classical regime $\hbar_\text{eff}\ll 1$. In this final section, we investigate how the localization length behaves as one crosses over into the deep quantum regime $\hbar_\text{eff}\gg 1$. To characterize this crossover, one may consider either the localization length defined from the average transmission, $\chi = -L/\ln\ev{T}$, or the one defined from the average logarithmic transmission, $\xi = -L/\ev{\ln T}$. As shown in the previous sections, $\chi$ is the quantity that properly interpolates to the critical cluster size in the limit $\hbar_\text{eff}\to 0$. However, it is also known that $\chi$ is not an appropriate measure of localization in the quantum regime, due to strong transmission fluctuations. In that regime, the relevant quantity is $\xi$, since it is constructed from $\langle\ln T\rangle$ which is self-averaging. Put differently, when $\hbar_\text{eff}\ll 1$, transport is dominated by free propagation within clusters and is thus physically captured by $\chi$ (see Sec.~\ref{Sec:logtransmission}), whereas in the quantum regime $\hbar_\text{eff}\gg 1$ transport is dominated by tunneling which is more faithfully captured by $\xi$. To describe the crossover between these two opposite regimes, we make the—admittedly somewhat arbitrary—choice to focus on $\xi$, as it is well defined in the deep quantum limit while still exhibiting, like $\chi$, a divergence at the percolation threshold in the semi-classical regime [Eq. (\ref{Eq:SC_asymptotics})].

\subsection{Numerical simulations}
\vspace{0.3cm}

\begin{figure}
 \centering       \includegraphics[width=0.7\textwidth]{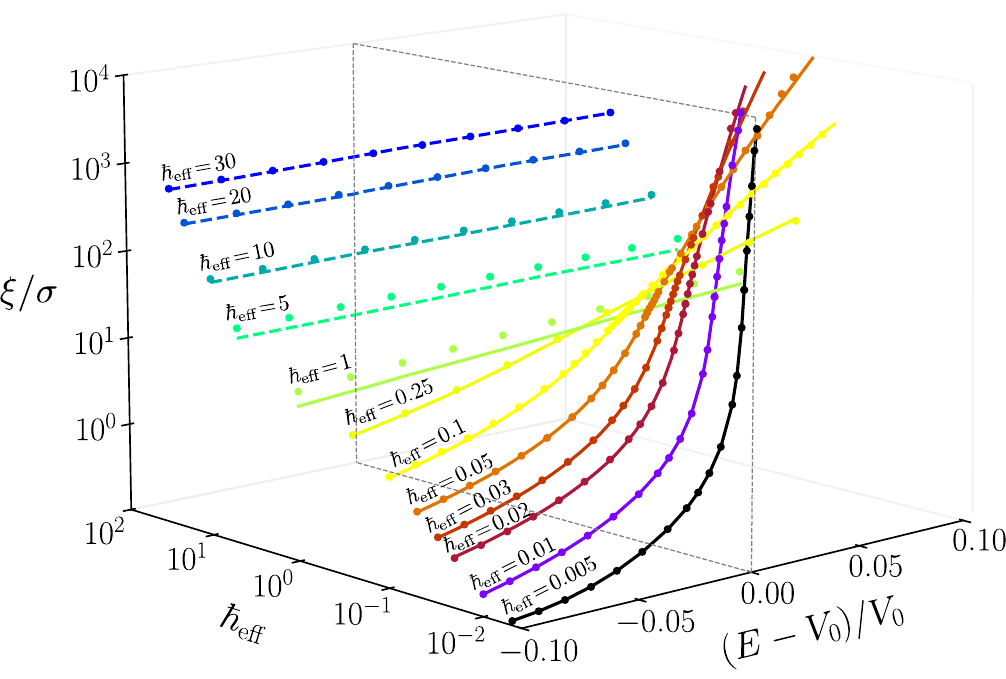}
 \caption{
 Localization length $\xi$ as a function of energy in the vicinity of the percolation threshold, for a wide range of $\hbar_\text{eff}$ values, probing the crossover from the semi-classical ($\hbar_\text{eff}\ll 1$) to the quantum regime ($\hbar_\text{eff}\gg 1$).
 Discrete point show $\xi$ extracted from transfer-matrix simulations, obtained from linear fits of $\langle \ln T\rangle=-L/\xi$ after averaging over $10^4$ disorder realizations.
 For each value of $\hbar_\text{eff}$,  the system size $L$ is chosen large enough to exceed $\xi$ while avoiding transmission values that fall below machine precision: $L/\sigma=10^1-10^2$ for $\hbar_\text{eff}\leq5$, $L/\sigma=10^2-10^3$ for $\hbar_\text{eff}=10$, and $L/\sigma=10^3-10^4$ for $\hbar_\text{eff}\geq20$. 
 Solid lines show the theoretical predictions (\ref{eq:xim1_SC}) in the semi-classical regime (plotted for $\hbar_\text{eff}\leq 1$). The dashed curve is the perturbative calculation (\ref{eq:xideepq}) in the deep quantum regime (plotted for $\hbar_\text{eff}\geq 5$).
 }
\label{Fig:xi_quantum_to_classical}
\end{figure}

We show in Fig.~\ref{Fig:xi_quantum_to_classical} the localization length $\xi$ as a function of energy in the vicinity of $E=V_0$, for a broad range of $\hbar_\text{eff}$ values ranging from $\hbar_\text{eff}\ll1$ to $\hbar_\text{eff}\gg1$. The colored points corresponds to results of transfer-matrix numerical simulations, where we use the same methodology as described in Sec.~\ref{sec:transfermatrix}, but with a finer spatial discretization $a=\hbar_\text{eff}\sigma/100$ for the data at $\hbar_\text{eff}>1$ to properly resolve the rapid oscillations of the wave function in this regime. 

As seen in Fig.~\ref{Fig:xi_quantum_to_classical}, when $\hbar_\text{eff}\ll1$ the localization length exhibits the expected divergence at the percolation threshold when $E\to V_0^-$. In this regime, the semi-classical prediction~\eqref{eq:xim1_SC} for $\xi$, shown as solid curves, agrees very well with the simulation results. In the strict classical limit $\hbar_\text{eff}=0$, $\xi$~becomes infinite for $E>V_0$ since the particle can then freely explore the entire system. Upon entering the quantum regime, $\hbar_\text{eff}\geq1$, quantum tunneling becomes significant and interference effects arising from multiple reflections between barriers lead to localization, regardless of whether the particle’s energy lies below or above the percolation threshold. The critical behavior associated with the percolation transition is therefore completely suppressed, and the localization length varies smoothly near $E\sim V_0$.

\subsection{Perturbation theory in the deep quantum regime}
\label{Sec:deep_quantum}
\vspace{0.3cm}

In the deep quantum regime, the simulation results can be compared with predictions from a fully quantum-mechanical perturbative treatment, where the disorder strength $V_0/E_\sigma \equiv 1/\hbar_\text{eff}^2 \ll 1$ serves as the small expansion parameter~\cite{Kuhn2007, Lugan2009}. Such a treatment can be carried out, for instance, using the phase formalism for Anderson localization developed in \cite{Lugan2009, Lugan2010} for speckle potentials. This method relies on the phase–amplitude representation $(\theta,r)$ of the wave function, $\psi(z)=r(z)\sin[\theta(z)], \quad \partial_z\psi(z)=kr(z)\cos(\theta(z))$, where for a weak disorder the phase can be expanded in powers of $V_0/E_\sigma$. This allows for a perturbation expansion of the Lyapunov exponent $\gamma\equiv\lim_{|x|\to\infty}\ev{\ln r(x)}/|x|$, which is related to the localization length  $\xi = -L/\ev{\ln T}$ controlling the transmission through $\xi=1/(2\gamma)$. At leading order, and for the disorder correlation function~\eqref{Eq:Vcorrelation}, one finds~\cite{Lugan2009}:
\begin{equation}
\label{eq:xideepq}
    \frac{1}{\xi(E)}\simeq  \frac{1}{\sigma}\frac{E_\sigma}{ E} \;\frac{1}{\hbar_\text{eff}^4} \;\int_{-\infty}^{0} \mathrm{d}u \; 
    \exp(-\frac{u^2}{2})\cos\left( 2u\sqrt{2{E}/{E_\sigma}} \right) + \mathcal{O}(\hbar_\text{eff}^{-6}).
\end{equation}
This relation is shown as the dashed curves in Fig.~\ref{Fig:xi_quantum_to_classical} for $\hbar_\text{eff}\geq5$. As expected, it agrees increasingly well with the numerical results as $\hbar_\text{eff}$ becomes larger.\newline

For the sake of clarity, we summarize in Table \ref{tab:reminder_table} the definitions, physical interpretations and regime of relevance of the two localization lengths~$\chi$ and $\xi$ introduced in the manuscript.
\begin{table}[h]
\centering
\begin{tabular}{c c c c}
\hline
Quantity & Definition & Physical meaning & Relevant regime \\
\hline\\[-0.2cm]
$\chi$ 
& $-[\rho\, \ln \ev{T}]^{-1}$
& Mean size of classically 
& Semi-classical regime \\
& 
&  allowed regions
& $\hbar_{\mathrm{eff}}\ll 1$ \\
[0.2cm]
$\xi$ 
& $-[\rho\, \ev{\ln T_b}]^{-1}$
& Exponential decay length
& Quantum regime \\
& 
& of quantum eigenstates
& $\hbar_{\mathrm{eff}}\gg 1$ \\
[0.2cm]
\hline
\end{tabular}
\caption{Summary of the definitions, physical interpretations, and regime of relevance of the two localization lengths introduced in the manuscript. $T_b$ is the transmission through a single barrier and $\rho$ is the density of barriers.}
\label{tab:reminder_table}
\end{table}

\section{Conclusion}
\vspace{0.3cm}

In this work, we investigated both numerically and theoretically the classical-to-quantum crossover between the percolation transition and Anderson localization in a red-detuned speckle potential. This crossover is controlled by an effective Planck constant, $\hbar_\text{eff}$, defined as the ratio between the de Broglie wavelength and the disorder correlation length. We found that, as the system departs from the classical limit $\hbar_\text{eff} = 0$, the characteristic algebraic divergence of the cluster length $\chi$ at the percolation threshold progressively evolves into a continuous localization length. For small but finite $\hbar_\text{eff}$, the correlated and non-Gaussian nature of the speckle potential plays a crucial role, causing the standard DMPK description of Anderson localization in uncorrelated disorder to break down. Using a semi-classical approach, we developed a consistent theoretical description of the localization length in this regime, correctly interpolating with the classical limit. Below the percolation threshold, we also identified the emergence of a bimodal transmission distribution, usually absent in 1D  models with uncorrelated Gaussian disorder, and which we associate with a diffusive motion of the particle quantum-mechanically transmitted and reflected near the potential maxima. We further provided an analytical derivation of this law within our semi-classical framework. Finally, we explored the global behavior of the localization length as $\hbar_\text{eff}$ is varied over several orders of magnitude, up to the deep quantum regime $\hbar_\text{eff} \gg 1$ where standard quantum perturbation theory becomes applicable.

Although Anderson localization was originally formulated for diagonal disorder with short-range correlations, its scope has been extended over the years to more general types of disorder. Our results provide an example of a situation in which Anderson localization in non-Gaussian, correlated disorder exhibits original yet non-universal properties, as already emphasized, e.g., in Refs.~\cite{Deych2003, Titov2005}. In the present case, properly accounting for the disorder statistics is essential to accurately capture the quantum-to-classical crossover. In particular, the correlated nature of the potential is crucial for the semi-classical regime~\eqref{eq:hbareff_def} to be well defined, even though we expect our results in the vicinity of the percolation threshold---most notably the bimodal distribution---to be generic and largely independent of the details of the short-range disorder correlation function owing to the universal character of the transition. In the 1D geometry considered here, the bounded nature of the potential is also essential for the existence of a percolation threshold, but this would not be the case in higher dimension. 

Our study also illustrates an interesting crossover between a regime featuring a genuine phase transition and another regime where no such transition exists. In disordered systems, similar situations arise, for instance, in dimensional crossovers from $d=3$ to $d<3$, or in crossovers between different symmetry classes such as in two-dimensional systems transitioning between the unitary and quantum-Hall classes \cite{DasSarma1997}, or between the unitary and orthogonal classes in three dimensions \cite{Hohenberg1977} where two distinct Anderson transitions occur.

The semi-classical regime explored in this work should be experimentally accessible in cold-atom experiments with optical speckle potentials, where the effective Planck constant can be tuned by adjusting the disorder amplitude. While earlier experiments \cite{Billy2008, Roati2008, Jendrzejewski2012, Semeghini2015} were deliberately designed to avoid classical trapping, with $\hbar_\text{eff}$ typically ranging from $0.5$ to $6$, the limit $\hbar_\text{eff}\ll1$ should be achievable by employing larger disorder amplitudes. Moreover, experiments now offer the possibility to precisely control the energy of atoms loaded into disordered potentials \cite{Volchkov2018}, opening  the way to a precise characterization of the critical behavior of  $\chi(E)$.

To deepen our understanding of the connection between percolation and Anderson localization, it would be  worthwhile to investigate whether a generalized 1D DMPK framework can be derived for non-Gaussian correlated disorder, capturing the change of the transmission distribution from log-normal to bimodal observed in the present work. This might be done, for instance, using the approach of Ref.~\cite{Gaspard2025}. It would also be highly valuable to explore the three-dimensional case, where a classical percolation transition exists while an Anderson transition occurs deep in the quantum regime. How these two transitions emerge or disappear as $\hbar_\text{eff}$ is increased remains unclear. Finally, probing the dynamics of Anderson localization could provide an original perspective on the percolation–Anderson crossover, for instance through the expansion of wave packets or the motion of their center-of-mass via the quantum boomerang effect~\cite{Prat2019, Tessieri2021}.

\section*{Appendix: Classical transmission distribution}
\vspace{0.3cm}
\label{AppendixA}

In this appendix, we derive the transmission distribution $P(T)$ in the classical limit, Eqs.~\eqref{eq:ptclassical} and \eqref{eq:doubledelta} of the main text. This distribution is defined as
\begin{equation}
\label{eq:PTclassical1}
    P(T)=\int_{-\infty}^\infty dt\,e^{-it T}\varphi_T(t),
\end{equation}
where $\varphi_T(t)$ denotes the characteristic function of the total transmission $T=T_b^{N_b}$. Since the transmission coefficients  $T_b$  of individual barriers are independent, their moments factorize as $\langle T^{n}\rangle=\langle T_b^n\rangle^{N_b}$. Thus the characteristic function becomes
\begin{equation}
\label{eq:varphiclassical}
    \varphi_T(t)\equiv\langle e^{itT}\rangle=
    \sum_{n=0}^\infty \frac{(it)^n}{n!}\langle T_b^n\rangle^{N_b}.
\end{equation}
When $E\geq V_0$, it is clear 
from the relation $T_b=\Theta(E-V)$ that $\langle T_b^n\rangle=1$ for all $n$, such that $\varphi_T(t)=\exp(it)$. Inserting this into Eq. (\ref{eq:PTclassical1}) immediately yields Eq. (\ref{eq:ptclassical}) of the main text. When $E\leq V_0$, we have:
\begin{equation}
    \langle T_b^n\rangle=
    \int_{0}^\infty d\omega\int_{-\infty}^{V_0} dV
    P(V,\omega)\Theta(E-V)^n=
    \int_{0}^\infty d\omega\int_{-\infty}^E dV
    P(V,\omega)\equiv \langle T_b\rangle,
\end{equation}
i.e., the $n$-dependence drops out so that all moments are equal. Equation~\eqref{eq:varphiclassical} then becomes 
\begin{equation}
    \varphi_T(t)=
    1+\sum_{n=1}^\infty \frac{(it)^n}{n!}\langle T_b\rangle^{N_b}=
    1+\langle T_\text{class}\rangle(e^{it}-1),
\end{equation}
where we used that $\langle T_b\rangle^{N_b}\equiv \langle T_\text{class}\rangle$. Substitution into Eq.~\eqref{eq:PTclassical1} directly leads to Eq.~(\ref{eq:doubledelta}) of the main text.

\ack{
The authors are indebted to  Lucile Julien,  Romain Pierrat and David Gaspard for  advice and useful discussions.}

\funding{This work was supported by the project Localization of Waves of the Simons Foundation (Grant No. 1027116, M.F.), and by Agence Nationale de la Recherche under Grant No ANR-23-PETQ-0001 Dyn1D France 2030.}
 

\roles{
M. Vrech and J. Major performed the numerical simulations and theoretical calculations, and contributed to writing the manuscript. D. Delande supervised the initial stage of the project, and M. Filoche supervised its subsequent development and contributed to writing the manuscript. N. Cherroret supervised the project, carried out theoretical calculations, and wrote the manuscript. }

\data{All numerical data presented in this work are available under reasonable request.}



\begin{thebibliography}{99}
\vspace{0.5cm}
\small



\bibitem{Anderson1958}
P. W. Anderson,
\emph{Absence of Diffusion in Certain Random Lattices},
Phys. Rev. \textbf{109}, 1492 (1958).

\bibitem{Lee1985}
P. A. Lee and T. V. Ramakrishnan, 
\emph{Disordered electronic systems},
Rev. Mod. Phys. \textbf{57}, 287 (1985).

\bibitem{Aspect2009}
A. Aspect and M. Inguscio, 
\emph{Anderson localization of ultracold atoms}, 
Phys. Today \textbf{62}, 30 (2009).

\bibitem{Cherroret2021}
N. Cherroret, T. Scoquart, and N. Cherroret,
\emph{Coherent multiple scattering of out-of-equilibrium interacting Bose gases},
Annals of Physics \textbf{435}, 168543 (2021).

\bibitem{Shapiro2012}
B. Shapiro, 
\emph{Cold atoms in the presence of disorder}, 
J. Phys. A: Math. Theor. \textbf{45}, 143001 (2012).

\bibitem{Hainaut2018}
C. Hainaut, I. Manai, J.-F. Cl\'ement, J. C. Garreau, P. Szriftgiser, G. Lemari\'e, N. Cherroret, D. Delande, and R. Chicireanu,
\emph{Controlling symmetry and localization with an artificial gauge field in a disordered quantum system},
Nature Comm. \textbf{9}, 1382 (2018).

\bibitem{An2018}
F. A. An, E. J. Meier, and B. Gadway,
\emph{Engineering a flux-dependent mobility edge in disordered zigzag chains},
Phys. Rev. X \textbf{8}, 031045 (2018).

\bibitem{Abanin2019}
D. A. Abanin, E. Altman, I. Bloch, and M. Serbyn,
\emph{Many-body localization, thermalization, and entanglement},
Rev. Mod. Phys. \textbf{91}, 021001 (2019).

\bibitem{Billy2008}
J. Billy, V. Josse, Z. Zuo, A. Bernard, B. Hambrecht,
P. Lugan, D. Cl\'ement, L. Sanchez-Palencia, P. Bouyer,
and A. Aspect, 
\emph{Direct observation of Anderson localization of matter waves in a controlled disorder}, 
Nature (London) \textbf{453}, 891 (2008).

\bibitem{Roati2008}
G. Roati, C. d’Errico, L. Fallani, M. Fattori, C. Fort, M. Zaccanti, G. Modugno, M. Modugno, and M. Inguscio, 
\emph{Anderson localization of a non-interacting Bose–Einstein condensate},
Nature (London) \textbf{453}, 895 (2008).

\bibitem{Jendrzejewski2012}
F. Jendrzejewski, A. Bernard, K. Muller, P. Cheinet,
V. Josse, M. Piraud, L. Pezz\'e, L. Sanchez-Palencia, A. Aspect, and P. Bouyer, 
\emph{Three-dimensional localization of ultracold atoms in an optical disordered potential},
Nat. Phys. \textbf{8}, 398 (2012). 

\bibitem{Semeghini2015}
G. Semeghini, M. Landini, P. Castilho, S. Roy, G. Spagnolli, A. Trenkwalder, M. Fattori, M. Inguscio, and
G. Modugno,
\emph{Measurement of the mobility edge for 3D Anderson localization},
Nature Physics \textbf{11}, 554 (2015).

\bibitem{Moore1995}
F. L. Moore, J. C. Robinson, C. F. Bharucha, B. Sundaram, and M. G. Raizen,
\emph{Atom Optics Realization of the Quantum $\delta$-Kicked Rotor}, 
Phys. Rev. Lett. \textbf{75}, 4598 (1995)

\bibitem{Chabe2008}
J. Chab\'e, G. Lemari\'e, B. Gr\'emaud, D. Delande, P. Szriftgiser, and J. C. Garreau,
\emph{Experimental Observation of the Anderson Metal-Insulator Transition with Atomic Matter Waves},
Phys. Rev. Lett. \textbf{101}, 255702 (2008).


\bibitem{Lugan2009}
P. Lugan, A. Aspect, L. Sanchez-Palencia, D. Delande, B. Gr\'emaud, C. A. M\"uller, and C. Miniatura,
\emph{One-dimensional Anderson localization in certain correlated random potentials},
Phys. Rev. A \textbf{80}, 023605 (2009) ; ibid,  Phys. Rev. A \textbf{84}, 019902 (2011).

\bibitem{Izrailev2012}
F. M. Izrailev, A. A. Krokhin, and N. M. Makarov,
\emph{Anomalous localization in low-dimensional systems with correlated disorder},
Phys. Reports \textbf{512}, 125 (2012).

\bibitem{Piraud2013}
M. Piraud and L. Sanchez-Palencia,
\emph{Tailoring Anderson localization by disorder correlations in 1D speckle potentials}, 
Eur. Phys. J. Special Topics \textbf{217}, 91 (2013).

\bibitem{Sanchez-Palencia2007}
L. Sanchez-Palencia, D. Cl\'ement, P. Lugan, P. Bouyer, G. V. Shlyapnikov, and A. Aspect,
\emph{Anderson Localization of Expanding Bose-Einstein Condensates in Random Potentials}, 
Phys. Rev. Lett. \textbf{98}, 210401 (2007).

\bibitem{Skipetrov2008}
S. E. Skipetrov, A. Minguzzi, B. A. van Tiggelen, and B. Shapiro,
\emph{Anderson Localization of a Bose-Einstein Condensate in a 3D Random Potential},
Phys. Rev. Lett. \textbf{100}, 165301 (2008).


\bibitem{Cherroret2012}
N. Cherroret, T. Karpiuk, C. A. M\"uller, B. Gr\'emaud, and C. Miniatura, 
\emph{Coherent backscattering of
ultracold matter waves : momentum space signatures}, Phys. Rev. A \textbf{85}, 011604(R) (2012)

\bibitem{Karpiuk2012}
T. Karpiuk, N. Cherroret, K. L. Lee, B. Gr\'emaud, C. A. Müller, C. Miniatura, 
\emph{Coherent forward
scattering peak induced by Anderson localization}, 
Phys. Rev. Lett. \textbf{109}, 190601 (2012).

\bibitem{Ghosh2014}
S. Ghosh, N. Cherroret, B. Gr\'emaud, C. Miniatura, D. Delande, 
\emph{Coherent forward scattering in
2D disordered systems}, 
Phys. Rev. A \textbf{90}, 063602 (2014)

\bibitem{Lee2014}
K. L. Lee, B. Gr\'emaud, C. Miniatura,
\emph{Dynamics of localized waves in 1D random potentials: statistical theory of the coherent forward scattering peak},
Phys. Rev. A \textbf{90}, 043605 (2014).

\bibitem{Kuhn2007}
R. C. Kuhn, O. Sigwarth, C. Miniatura, D. Delande, and C. A. M\"uller,
\emph{Coherent matter wave transport in speckle potentials},
New. J. Phys. \textbf{9}, 161 (2007).

\bibitem{Yedjour2010}
A. Yedjour and B. A. Tiggelen, 
\emph{Diffusion and localization of cold atoms in 3D optical speckle},
Eur. Phys. J. D \textbf{59}, 249 (2010).

\bibitem{Delande2014}
D. Delande, G. Orso,
\emph{Mobility Edge for Cold Atoms in Laser Speckle Potentials},
Phys. Rev. Lett. \textbf{113}, 060601 (2014).

\bibitem{Pasek2017}
M. Pasek, G. Orso, and D. Delande,
\emph{Anderson localization of ultracold atoms: Where is the mobility edge?},
Phys. Rev. Lett. \textbf{118}, 170403 (2017).

\bibitem{Casati1990}
G. Casati, I. Guarneri and D. Shepelyansky, 
\emph{Classical chaos, quantum localization and fluctuations: A unified view},
Physica A \textbf{163}, 205 (1990).

\bibitem{Biddle2010}
J. Biddle and S. Das Sarma, 
\emph{Predicted Mobility Edges in One-Dimensional Incommensurate Optical Lattices: An Exactly Solvable Model of Anderson Localization}, 
Phys. Rev. Lett. \textbf{104}, 070601 (2010).

\bibitem{Ganeshan2015}
S. Ganeshan, J. H. Pixley, and S. Das Sarma, 
\emph{Nearest Neighbor Tight Binding Models with an Exact Mobility Edge in One Dimension}, 
Phys. Rev. Lett. \textbf{114}, 146601 (2015).

\bibitem{Wang2020}
Y. Wang, X. Xia, L. Zhang, H. Yao, S. Chen, J. You,
Q. Zhou, and X.-J. Liu, 
\emph{One-Dimensional Quasiperiodic Mosaic Lattice with Exact Mobility Edges},
Phys. Rev. Lett. \textbf{125}, 196604 (2020).

\bibitem{Vynck2023}
K. Vynck, R. Pierrat, R. Carminati, L. S. Froufe-P\'erez, F. Scheffold, R. Sapienza, S. Vignolini, J. Jos\'e S\'aenz,
\emph{Light in correlated disordered media},
Rev. Mod. Phys. \textbf{95}, 045003 (2023).

\bibitem{Kirkpatrick1973}
S. Kirkpatrick,
\emph{Percolation and conduction},
Rev. Mod. Phys. \textbf{54}, 574 (1973).

\bibitem{Isichenko1992}
M. B. Isichenko,
\emph{Percolation, statistical topography, and transport in random media},
Rev. Mod. Phys.  \textbf{64}, 961 (1992).

\bibitem{Saberi2015}
A. A. Saberi,
\emph{Recent advances in percolation theory and its applications},
Phys. Reports \textbf{578}, 1 (2015).

\bibitem{Pezze2011}
L. Pezz\'e, M. Robert-de-Saint-Vincent, T. Bourdel, J.-P. Brantut, B. Allard, T. Plisson, A. Aspect, P. Bouyer and L. Sanchez-Palencia,
\emph{Regimes of classical transport of cold gases in a
two-dimensional anisotropic disorder},
New Journal of Physics \textbf{13} (2011) 095015.

\bibitem{Morong2015}
W. Morong and B. DeMarco,
Simulation of Anderson localization in two-dimensional ultracold gases for pointlike disorder,
Phys. Rev. A \textbf{92}, 023625 (2015).

\bibitem{Abrahams1979}
E. Abrahams, P. W. Anderson, D. C. Licciardello, and T. V. Ramakrishnan,
\emph{Scaling Theory of Localization: Absence of Quantum Diffusion in Two Dimensions},
Phys. Rev. Lett. \textbf{42}, 673 (1979).

\bibitem{Filoche2024}
M. Filoche, P. Pelletier, D. Delande, and S. Mayboroda,
\emph{Anderson mobility edge as a percolation transition},
Phys. Rev. B \textbf{109}, L220202 (2024).



\bibitem{Prat2016}
T. Prat, N. Cherroret, and D. Delande,
Semiclassical spectral function and density of states in speckle potentials,
Phys. Rev. A \textbf{94}, 022114 (2016).

\bibitem{Trappe2015}
M. I. Trappe, D. Delande and C. A. Müller, 
\emph{Semiclassical spectral function for matter waves in random potentials},
J. Phys. A: Math. Theor. \textbf{48}, 245102 (2015).

\bibitem{Abrikosov1981}
A. A. Abrikosov, 
\emph{The paradox with the static conductivity of a one-dimensional metal}, 
Solid State Comm. \textbf{37}, 997 (1981).

\bibitem{Mueller2011}
C. A. M\"uller and D. Delande,
\emph{Disorder and interference: localization phenomena}, 
in "Les Houches 2009 - Session XCI: Ultracold Gases and Quantum Information" edited by C. Miniatura et al. (Oxford University Press, 2011).

\bibitem{Goodman2007}
J. W. Goodman, 
\emph{Speckle phenomena in optics: theory
and applications} (Roberts and Company, 2007).

\bibitem{Clement2006}
D. Cl\'ement, A. F. Var\'on, J. A. Retter, L. Sanchez-Palencia, A. Aspect, and P. Bouyer, 
\emph{Experimental study of the transport of coherent interacting matter-waves in a 1D random potential induced by laser speckle},
New Journal of Physics  \textbf{8}, 165 (2006).

\bibitem{Holstein1982}
B. R. Holstein and A. R. Swift, 
\emph{Path integrals and the WKB approximation}, 
American Journal of Physics \textbf{50}, 829 (1982).

\bibitem{Maitra1996}
N. T. Maitra and E. J. Heller,
\emph{Semiclassical perturbation approach to quantum reflection}, 
Phys. Rev. A \textbf{54}, 4763 (1996).

\bibitem{Beenakker1997}
C. W. J. Beenakker,
\emph{Random-matrix theory of quantum transport},
Rev. Mod. Phys. \textbf{69}, 731 (1997).

\bibitem{Dettmann2009}
C. P. Dettmann and O. Georgiou,
\emph{Product of $n$ independent uniform random variables},
Statistics and Probability Lett.
\textbf{79}, 2501 (2009).

\bibitem{Deych2003}
L. I. Deych, M. V. Erementchouk, and A. A. Lisyansky,
\emph{Scaling properties of 1D Anderson model with correlated diagonal disorder},
Physica B: Condensed Matter \textbf{338}, 79 (2003).

\bibitem{Titov2005}
M. Titov and H. Schomerus,
\emph{Nonuniversality of Anderson localization in short-range correlated disorder},
Phys. Rev. Lett. \textbf{95}, 126602 (2005).

\bibitem{DasSarma1997}
S. Das Sarma, in
\emph{Perspectives in Quantum Hall Effects}, 
Edited by S. Das Sarma and A. Pinczuk, John Wiley $\&$ Sons,1997, p. 1.

\bibitem{Hohenberg1977}
M. Batsch, L. Schweitzer, I. Kh. Zharekeshev, and B. Kramer 
 \emph{Crossover from critical orthogonal to critical unitary statistics at the Anderson transition}, 
Phys. Rev. Lett. \textbf{77}, 1552 (1996).

\bibitem{Gaspard2025}
D. Gaspard and A. Goetschy,
\emph{Radiant Field Theory: A Transport Approach to Shaped Wave Transmission through Disordered Media},
Phys. Rev. Lett. \textbf{135}, 033804 (2025).

\bibitem{Prat2019}
T. Prat, D. Delande, N. Cherroret, 
\emph{Quantum boomeranglike effect of wave packets in random potentials}, 
Phys. Rev. A \textbf{99}, 023629 (2019).

\bibitem{Tessieri2021}
L. Tessieri, Z. Akdeniz, N. Cherroret, D. Delande, P. Vignolo, 
\emph{Quantum boomerang effect : beyond
the standard Anderson model}, 
Phys. Rev. A \textbf{103}, 063316 (2021).

\bibitem{Lugan2010}
P. Lugan, 
\emph{Ultracold Bose gases in random potentials: collective excitations and localization effects},
PhD thesis, Ecole Polytechnique (2010).

\bibitem{Volchkov2018}
V. Volchkov, M. Pasek, V. Denechaud, M. Mukhtar, A. Aspect, D. Delande, V. Josse, 
\emph{Measurement of spectral functions of ultracold atoms in disordered potentials}, 
Phys. Rev. Lett. \textbf{120}, 060404 (2018).

\end{thebibliography}
\end{document}